# A 3D Multiscale Modelling of Cortical Bone Structure, Using the Inverse Identification Method: Microfibril Scale Study

**Abdelwahed Barkaoui,** Ridha Hambli
**Institute PRISME, EA4229, University of Orleans**
**Polytech' Orléans, 8, Rue Léonard de Vinci 45072 Orléans, France**
Abdelwahed.barkaoui@etu.univ-orleans.fr

## ABSTRACT

Complexity and heterogeneity of bone tissue require a multiscale modelling to understand their mechanical behaviour and their remodelling mechanism. Human cortical bone structure consists of six structural scale levels which are the (macroscopic) cortical bone, osteonal, lamellar, fibrous, fibril and microfibril. In this paper, a 3D model based on finite elements method was achieved to study the nanomechanical behaviour of collagen Microfibril. The mechanical properties and the geometry (gap, overlap and diameter) of both tropocollagen and mineral were taken into consideration as well as the effects of cross-links. An inverse identification method has been applied to determine equivalent averaged properties in order to link up these nanoscopic characteristics to the macroscopic mechanical behaviour of bone tissue. Results of nanostructure modelling of the nanomechanical properties of strain deformation under varying cross-links were investigated in this work.

## 1. Introduction

A long bone like the femur consists of three parties from the center outward: the marrow, the spongy bone and cortical bone. In this study we are interested only in compact bone. A microscopic analysis reveals a complex architecture that can be described as follows. The bone is a composite material: it must imagine hollow cylinders juxtaposed next to each other and sealed by a matrix. The cylinders are called Osteon, the inner bore Haversian canal and the matrix pore system. Further analysis shows that osteons are in fact an assembly of cylindrical strips embedded in each other and each blade is composed of a network of fibers wound helically oriented collagen and inserted into hydroxyapatite crystals. The orientation of collagen fibers may be different between two consecutive slices. These fibers are one set of fibrils. Each fibril is in turn composed of micro fibrils. Finally, each micro fibril is a helical arrangement of five tropocollagen molecules. Fig1.provides a better understanding of this large complex architecture.

In the previous multiscale studies of cortical bone, it's used to be started by microfibril as compound mentioning only its geometry and its special arrangement of molecular rows, with neither regarding less its mechanical behaviours, on other word there are neither analytical studies nor numerical modeling at this level of scale. The most notable studies are those of Jager and Fratzl, 2000, Andreas Fritsch 2009, Markus 2008, [1, 2, 3, 4]. However, this work focuses on modeling of fibril scale, that is to say a larger scale than the one we want, our job is to fill this gap.

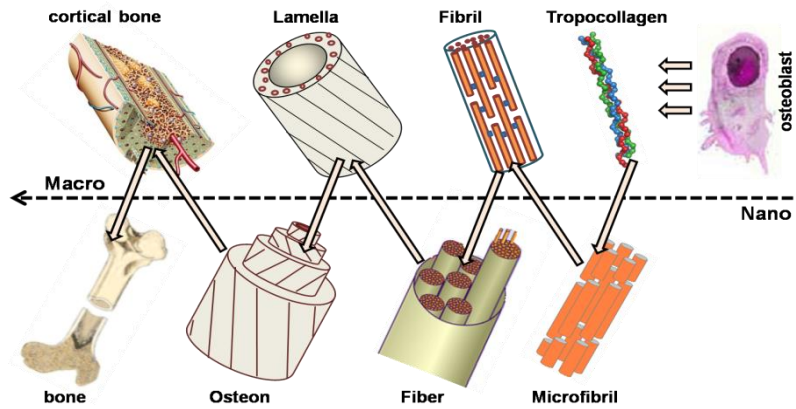

**Figure 1.**the hierarchical structure of cortical bone

## 2. Microfibril structure and composition

The microfibril is a helical assembly of five tropocollagen molecules (rotational symmetry of order 5), which are offset one another with an interval D (~ 67 nm) and creates a cylindrical formation with a diameter of about 4 nm. The orientation and axial arrangement of tropocollagen molecules in the microfibril have been deducted from an electron-microscopic observation showing transverse striations with a period D. The origin of this streaking was performed by a gradation in the arrangement of elements that are staggered tropocollagen with themselves at an interval D. [5, 6, 7, 8]. Fig2.

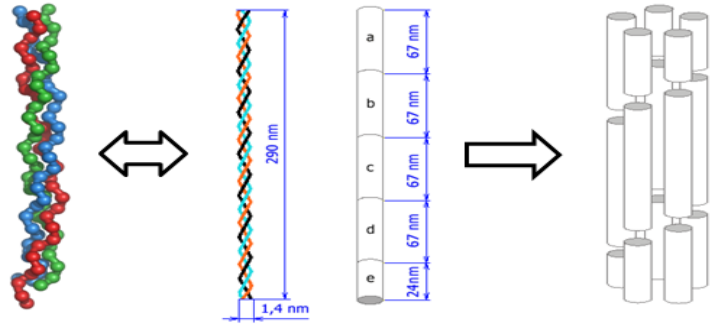

**Figure 2**.Schematic illustration of the three tropocollagen molecules and the formation of the microfibril

The mineral is almost entirely of crystalline hydroxyapatite $Ca_{10}(PO_4)_6(OH)_2$ impure. This happens, a mature form of needles or leaves, included into the gap and between the tropocollagen molecular. The size of the sheets is several tens or several hundreds of nm in their plan, and a few nm in thickness.

In the biological sciences, crosslinking typically refers to a more specific reaction used to probe molecular interactions. For example, proteins can be cross-linked together using small-molecule crosslinkers. In biological tissue, crosslinks can be induced as disulfide bonds between tropocollagen molecules. Compromised collagen in the cornea, a condition known as keratoconus, can be treated with clinical crosslinking. In our finite element model the cross-links are modeled as a spring with defined stiffness.

## 3. Methods

In this work we have used the method of finite elements in order to investigate the mechanical behavior of the microfibril, its damping capacity and its fracture resistance. The outputs of this simulation are used as inputs for the inverse identification method to identify equivalent properties, Young's modulus and Poisson's ratio of the microfibril.

A three- dimensional model of collagen microfibril with symmetric and periodic boundary conditions is considered here, with an array of 5 tropocollagen molecules cross-linked together using springs, the all is put into a mineral matrix. An entire plan of experience has been considered in order to investigate the influence of all geometric and mechanical parameters on the mechanical behaviour of the microfibril under the varying number of cross-links.

An inverse method was applied to identify the equivalent properties. A Newton Raphson algorithm written in python was coupled to Abaqus code, allows us to identify these properties.

## 4. Results

Below an example of the results of finite element simulation with E1 is the Young's modulus of collagen and E2 is the Young's modulus of the mineral. Fig4 illustrate the damping capacity of microfibril under different loading. Both graphs show the effect of number of cross links on the mechanical behaviour of the microfibril.

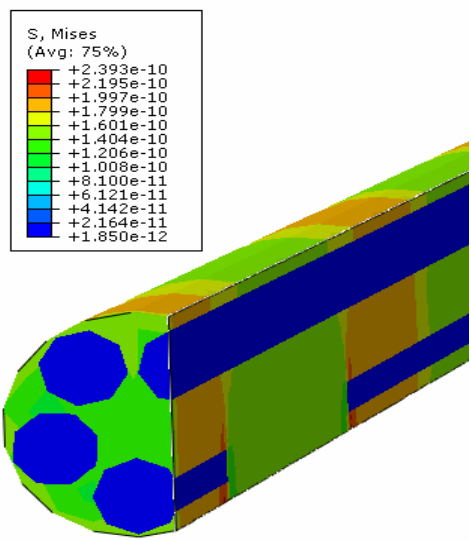
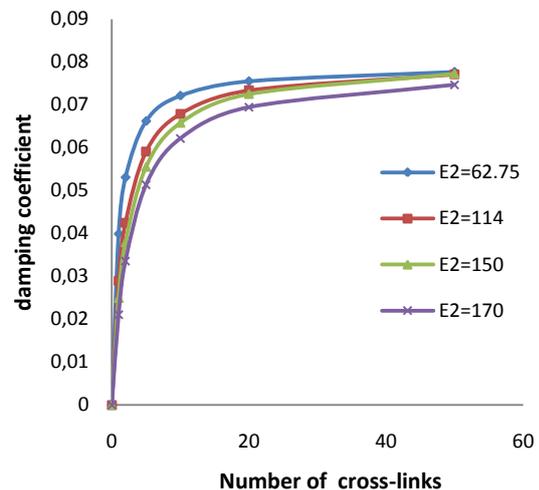

**Figure3.** FE Von Mises stress contour

**Figure 4.** Relation between damping coefficient and number of cross-links with E1=2GPa

The maximum efforts determined by finite element numerical simulation are introduced in the program of inverse identification method in order to indentify equivalent properties. The two graphs below illustrate some of the results.

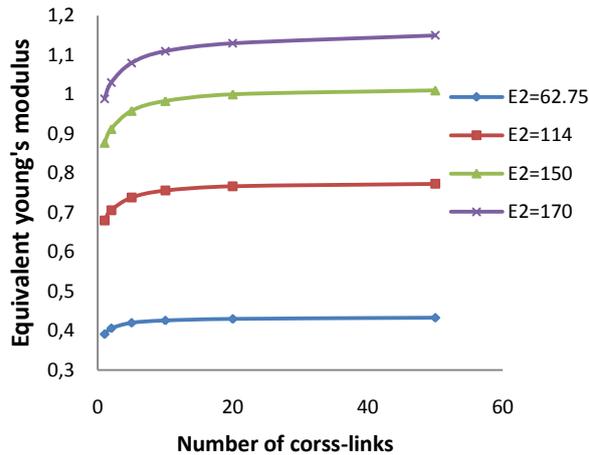 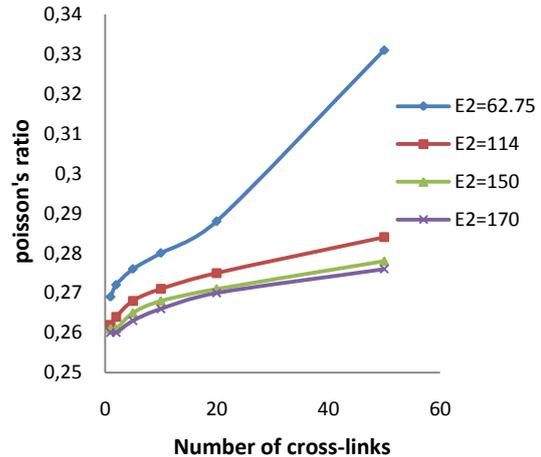

**Figure 5.** Relation between Young's modulus and number of cross-links with E1=2GPa

**Figure 6.** Relation between Poisson's ratios and number of cross-links with E1=2GPa

## 5. Conclusion

In this paper we study for the first time the mechanical behaviour of the microfibril. This work also allows us to understand better this nanoscale and study the upper level scale which is the collagen fibril with the same methods by using the results found in this scale.

**Acknowledgements**

This work has been supported by French National Research Agency (ANR) through TecSan program (Project MoDos, n°ANR-09-TECS-018).

**References**

[1] I. Jäger, P. Fratzl, Mineralized collagen fibrils: A mechanical model with a staggered arrangement of mineral particles, Biophys. J. 79 (2000) 1737–1746

[2] A.Fritsch, C.Hellimch, L.Dormieux, Ductile sliding between mineral crystals followed by rupture of collagen crosslinks: Experimentally supported micromechanical explanation of bone strength, J. Of theoretical biology 260(2009) 230-252.

[3] M.J.Buehler, Nanomechics of collagen fibrils under varying cross-links densities: Atomistic and continuum studies, J.of the mechanical behaviour of biomedical materials (2008) 59-67.

[4] M.J.Buehler, Molecular architecture of collagen fibrils: A critical length scale for tough fibrils, Current applied physics 8 (2008) 440-442.

[5] A. Blažej, A. Galatík, J. Galatík, M. Mládek, Technologie kůže a kožešin, SNTL Praha 1984

[6] P.Bruckner, D. E Birk, Collagen, 2005

[7] M. Stančíková, R. Stančík, Z. Gubzová, J. Rovenský, Collagen in the Treatment of Rheumatic Diseases - Oral Tolerance, Bratislavské listy 1999

[8] Stephen C. Cowin, Do liquid crystal-like flow processes occur in the supramolecular assembly of biological tissues, J. Non-Newtonion Fluid Mech. 119 (2004) 155-162.